\documentclass[runningheads]{llncs}
\usepackage{graphicx}
\usepackage{amsmath,amssymb,amsfonts}
\usepackage{booktabs} 
\usepackage{graphicx}
\usepackage{amsfonts}
\usepackage{booktabs}
\usepackage{todonotes}
\usepackage{subcaption}
\usepackage{enumitem}

\newcommand*{\inlineequation}[2][]{%
  \begingroup
    \refstepcounter{equation}%
    \ifx\\#1\\%
    \else
      \label{#1}%
    \fi
    \relpenalty=10000 %
    \binoppenalty=10000 %
    \ensuremath{%
      #2%
    }%
    ~\@eqnnum
  \endgroup
}
\makeatother

\usepackage[colorlinks=true,allcolors=blue]{hyperref}

\newcommand\nar[1]{}
\newcommand\alex[1]{}
\newcommand\sarah[1]{}
\newif\ifempirical\empiricalfalse

\begin{document}
\title{Decentralisation Conscious Players And System Reliability}
%
%

\author{%
    Sarah Azouvi\inst{1} \and
    Alexander Hicks\inst{2}
}
%
%
\institute{
      Protocol Labs\\
      \email{sarah.azouvi@protocol.ai}\and
      University College London\\
      \email{alexander.hicks@ucl.ac.uk}
}

\maketitle              
%

\newcommand\target{\mathsf{target}}
\newcommand\nonce{\mathsf{nonce}}
\newcommand{\blockdata}{\mathsf{blockdata}}
\newcommand\eq{\textit{eq}}
\newcommand\new{\text{new}}
\newcommand{\xeq}{x_\eq}

\begin{abstract}
We propose a game-theoretic model of the reliability of decentralised systems based on Varian's model of system reliability~\cite{varian2004system}, to which we add a new normalised total effort case that models \textit{decentralisation conscious players} who prioritise decentralisation. 

We derive the Nash equilibria in the normalised total effort game.
In these equilibria, either one or two values are played by players that do not free ride.
The speed at which players can adjust their contributions can determine how an equilibrium is reached and equilibrium values.
The behaviour of decentralisation conscious players is robust to deviations by other players.

Our results highlight the role that decentralisation conscious players can play in maintaining decentralisation.
They also highlight, however, that by supporting an equilibrium that requires an important contribution they cannot be expected to increase decentralisation as contributing the equilibrium value may still imply a loss for many players.
We also discuss practical constraints on decentralisation in the context of our model.
\end{abstract}

\section{Introduction}
The reliability of a system captures the likelihood that it performs as intended.
For a decentralised system, there are two important components to consider, the number of participants and the distribution of power between them~\cite{troncoso2017systematizing}.
Even if there is a high number of contributors, if one of them has significantly more control over the system, there will be no meaningful level of decentralisation.
This presents a problem that has been hard to solve in practice.
How can the effort put into a system grow while maintaining an acceptable level of decentralisation?

Participation rewards can incentivise an increase in the effort invested in a system but a greater total effort can also be more centralised.
Certain protocol considerations may alleviate this effect, e.g., at the consensus level~\cite{bano2019sok}.
It is also sometimes assumed that a portion of players will behave altruistically, following protocol guidelines even when an a priori more profitable strategies exist.

An alternative assumption, which we consider here, is that players have an incentive to maintain decentralisation.
Short-term profits may be outweighed by the possible long-term profits associated with maintaining a reliable system.
For example, the value of a cryptocurrency that is vulnerable to hostile takeovers may decrease so miners have an incentive to maintain decentralisation and preserve the value of the tokens they hold and continue to receive.

Three observations support this assumption.
First, the market price of a cryptocurrency is linked to its security~\cite{bissias2020pricing}.
Second, numerous flaws have been identified in the incentive structure of cryptocurrencies~\cite{bonneau2015sok,judmayer2019pay}, yet attacks based on these have scarcely been observed~\cite{neudecker2019short}.
Third, a mining pool has previously acted to avoid controlling more than half of Bitcoin's hash rate~\cite{hajdarbegovic_2014}.

To further understand the rationality of maintaining decentralisation, this paper studies a game-theoretic model of decentralisation conscious players who prioritise decentralisation.
With this model, we can analyse how such players will behave to ensure that a system remains decentralised, what effort they may contribute, and under which circumstances they will free-ride.

\paragraph{Our contributions}
The main contribution of this paper is the introduction and analysis of the normalised total effort game with decentralisation conscious players that extends Varian's system reliability model to decentralised systems.

We introduce our model based on the normalised total effort (NTE) function in the context of Varian's system reliability model~\cite{varian2004system} in Section~\ref{sec:background}.
In Section~\ref{sec:reliability}, we derive the two types of Nash equilibria between decentralisation conscious players in which players contribute the same amount or two distinct amounts while others free ride.
We also consider the social optimum, in which players contribute the same effort while minimising their costs to maximise decentralisation.

To understand how decentralisation conscious players will behave in real systems alongside selfish and Byzantine players, we study in Section~\ref{sec:robustness} the robustness of the previously derived equilibria when (i) the number of players change, which does not always affect the equilibrium; (ii) players deviate from the equilibrium, which can lead to a new equilibrium where players (possibly fewer) contribute a greater effort.
Non-myopic players may, therefore, be incentivised to deviate from an equilibrium to reach a new equilibrium with fewer contributing players and a greater share of rewards.

Finally, we discuss in Section~\ref{sec:discussion} some practical constraints on decentralisation in relation to our model.

\section{Modelling System Reliability And Normalised Total Effort}\label{sec:background}
Varian's original model of system reliability (treated as a public good) considers three cases based on how the individual efforts $x_i$ of players are factored in~\cite{varian2004system}.
The weakest link case considers the minimal effort exerted by any one of the players i.e, $F(x_1,\ldots,x_n)=\min_{i}(x_i)$.
The total effort case considers the sum of every player's efforts i.e., $F(x_1,\ldots,x_n)=\sum_{i=1}^{n}x_i$.
The best shot case considers the maximal effort exerted by any one of the players i.e., $F(x_1,\ldots,x_n)=\max_{i}(x_i)$.

Reliability will usually depend on a combination of these cases.
For example, in the case of software security, a program's correctness can depend on the weakest link (the developer that introduces bugs), vulnerability testing depends on the total effort of all the testers, and the contributions of a system architect maps to the best shot case~\cite{anderson2020security}.

For each case, the Nash equilibria can be computed with the expected pay-off $u_i$ for a player $i$ expressed as in Equation~\ref{eq:utility_varian}, as can be the social optimum based on social pay-off $SP$ expressed as in Equation~\ref{eq:varian_social_payoff}.
The likelihood that the system operates successfully is captured by $P(F(x_1,\ldots,x_n))$, which is assumed to be differentiable, increasing, and concave.
The parameter $v_i$ is the value derived by player $i$ of the system operating successfully, and $c_i x_i$ is the cost to player $i$ where $c_i$ is a constant.
The choice of a linear cost function of the form $c_ix_i$ implicitly ignores more complex forms of cost and any fixed costs.
This is a limitation but it is realistic in relevant cases e.g., the energy required to operate a computer may be valued at a fixed price per kilowatt-hours.

\begin{align}
& u_i = P(F(x_1,\ldots,x_n))v_i- c_ix_i \label{eq:utility_varian} \\
& SP=P(F(x_1,\ldots,x_n))\sum_{i=1}^{n}v_i-\sum_{i=1}^nc_ix_i \label{eq:varian_social_payoff}
\end{align}

The equilibria can be used to determine when free-riding can be expected to occur based on the form of $F$.
For example, in the total effort case, the equilibrium is for players to free ride on the player who has the highest benefit-cost ratio $\frac{v_i}{c_i}$.
The social optimum, obtained by maximising the social pay-off rather than the player's utility functions, can also reveal how selfish behaviour from the players will lead to an outcome that is different from the social optimum.
This is the case in the total effort case used as an example.
Players free ride on the player with the highest benefit-cost ratio, which amounts to less total effort than in the social optimum, and the ``wrong'' players (those with the smallest benefit-cost ratio) can be found to contribute that effort.

The takeaway from Varian's results is that centralisation emerges even in the total effort case that involves everyone's contributions, and that rational behaviour can conflict with the social optimum i.e., selfish behaviour can lead to a weaker system -- a concept known as \emph{the price of anarchy}~\cite{roughgarden2005selfish}.
If decentralisation is desired, this means that an alternative model that produces individual and social outcomes that support a decentralised and stronger system is required.

To model decentralisation, the \textit{relative} contribution of every player in the system must be taken into account because while the total effort should be as high as possible, the effort must also be as evenly distributed as possible.
In practice, however, there are trade-offs between maximising total effort and distributing effort evenly.
It is unlikely that every player will have the same capacity to contribute, so maximising the total effort is likely to come at the cost of a uniform distribution of effort, and vice versa.

With this in mind, we define in Equation~\ref{eq:normeffort} the normalised total effort (NTE) function based on the total effort and the maximal contribution.
If the total effort is high but the maximal effort is also high then the NTE may not be as high as when the total effort is high but the maximal effort is low.

\begin{equation} \label{eq:normeffort}
F(x_1,\ldots,x_n) = \frac{\sum_{i=1}^{n} x_i}{\max_i(x_i)} 
\end{equation}

The normalised total effort function is scale invariant i.e., $F(\alpha x_1, \ldots, \alpha x_n) = F(x_1,\ldots,x_n)$ for any $\alpha$.
This is because we are modelling players who care about decentralisation over total effort.
The goal is to capture the fact that in systems that are designed to be decentralised, it is not only the total effort (studied by Varian) that matters but the distribution of effort and, in particular, how much the maximal contribution by a single player is as a portion of the total effort, which our measure captures.
Finding a measure that captures both this and the benefits of a higher total effort is an open problem, and measures similar to ours (e.g., the work of Kwon et al.~\cite{Kwon2019impossibility}) suffer from the same limitation.

We show in Section~\ref{sec:robustness} that contributions can still be expected to increase given that other players who prioritise maximising their share of rewards exist.
Thus, much like in software security, a decentralised system's reliability depends on nodes that are primarily concerned with decentralisation (normalised total effort) and nodes that are primarily concerned with higher contributions (and higher rewards) that increase the best shot and total effort. 
Because the best shot and total effort case have already been studied by Varian, our focus in this paper is the normalised total effort case.

\section{Equilibria Between Decentralisation Conscious Players}\label{sec:reliability}\label{sec:NE}
We begin by studying the Nash equilibria of the NTE game defined below.

\begin{definition}[Normalised Total Effort Game]\label{def:game}
We call the normalised total effort game (NTEG) the game consisting of $n$ players with costs $(c_1,\ldots,c_n)\in(\mathbb{R}^*_+)^n$, valuations $(v_1,\ldots,v_n)\in(\mathbb{R}^*_+)^n$, contributions $(x_1,\ldots,x_n)\in(\mathbb{R}_+)^n$, utility functions defined by equations~\ref{eq:normeffort} and~\ref{eq:utility_varian}, benefit-cost ratios $\beta_i=\frac{v_i}{c_i}$ such that $\beta_1<\ldots<\beta_n$, and where we assume a logarithmic reliability function $P(F(x_1,\dots,x_n))=\ln\big(\frac{\sum_{i=1}^{n} x_i}{\max_i(x_i)}\big)$ for $\max_i(x_i)>0$.
By convention we have $P(F(0,\dots,0))=0$ i.e., a system with no contributions does not function.
\begin{align}
& F(x_1,\ldots,x_n) = \frac{\sum_{i=1}^{n} x_i}{\max_i(x_i)} \tag{\ref{eq:normeffort}} \\
& u_i = v_iP(F(x_1,\ldots,x_n))- c_i x_i \tag{\ref{eq:utility_varian}} 
\end{align}
\end{definition}

\paragraph*{Two-player case}
We start by considering the simple case of a two-player game and the following theorem, which we prove in Appendix~\ref{proof:two-player-ne}.

\begin{theorem}\label{thm:two-player-ne}
In a two-player NTEG, the Nash equilibria are for both players to contribute the same effort $x_1 = x_2 =\xeq$
such that $\xeq \le \frac{1}{2}\min(\beta_1,\beta_2)$.
\end{theorem}

Both players contribute the same effort when the equilibrium is played, which is the only possible ``decentralised'' solution.

\paragraph*{Multiplayer case}
For $n>2$ players, we prove the following in Appendix~\ref{proof:nash-equilibria}.

\begin{theorem}\label{thm:nash-equilibria}
In a $n>2$ player NTEG, there exist two types of equilibrium.
\begin{enumerate}
\item ($1$-value equilibria) players $i+1$ to $n$ (for $1\le i < n$) contribute $\xeq$ subject to the constraint expressed by Inequality~\ref{eq:multiplayerNE} and players 1 to $i$ with the smallest benefit-cost ratio free ride on them.

\begin{equation}\label{eq:multiplayerNE}
\frac{1}{n-i}{\beta_{i}}\le \xeq \le\frac{1}{n-i}\beta_{i+1}
\end{equation}

\item ($2$-value equilibria) player $i$
contributes $x_m$ , players ($i+1$ to $n$) contribute $x_M$, where $x_m<x_M$, subject to the constraints in Inequality~\ref{eq:2-value-eq-cond1} and Equation~\ref{eq:2-value-eq-cond2} and players 1 to $i-1$ free ride, for $1\le i\le n$ (with no players free riding if $i=1$).
\end{enumerate}

\begin{align}
& \frac{1}{n-i+1}{\beta_{i}}<x_M<\frac{1}{n-i}\beta_{i}  \label{eq:2-value-eq-cond1}  \\
& x_m = \beta_i - (n-i)x_M \label{eq:2-value-eq-cond2}
\end{align}
\end{theorem}

We highlight Lemma~\ref{cor:maxtwocontrib} (proven as part of the proof) that we will reuse later.

\begin{lemma}\label{cor:maxtwocontrib}
If there exist two contributing rational players whose contributions are strictly less than $\max_{i}(x_i)$ and who play their best strategy, then those players must have the same benefit-cost ratio.
\end{lemma}

Unless specified otherwise, we denote by $\xeq$ the value played by the players or bulk of players in the 1-value or 2-value equilibrium, respectively.
For both types of equilibrium, the lower $\xeq$ is the more decentralised the system is, as more players can contribute and the less free-riding there is.

The fact that one equilibrium is for all players to contribute the same amount of effort makes sense as the NTE function encodes the social goal of maximising decentralisation.
It also prevents the perverse effects of any feedback loops that enable some players to contribute increasingly more than other players.

The 2-value equilibrium is less expected.
It shows that, even if some players cannot match the other players' contributions (due to their own costs or valuation), they may still be incentivised to contribute.

\subsection{The impact of a reward}
The equilibria we have derived above include the case where everyone contributes no effort.
Adding a reward function $R_i(x_1,\dots,x_n)$ to the utility function, as in Equation~\ref{eq:reward}. (e.g., cryptocurrency mining rewards) is a way of explicitly incentivising non-zero contributions, particularly from new players.

\begin{equation}\label{eq:reward}
u_i = P(F(x_1,\dots,x_n))v_i- c_ix_i + R_i(x_1,\dots,x_n)
\end{equation}

A reward separate from the valuation $v$ models the compensation for the effort invested in the system rather than the benefit derived from being able to use the system.
In practice, it may be a constant $R$ that can be won by players with a probability proportional to the effort they contribute.
Under certain conditions, this is an optimal allocation rule~\cite{chen2019axiomatic} so we restrict ourselves to this case.

\begin{equation}\label{eq:propreward}
    R_i(x_1,\dots,x_n)= \left\{ \begin{array}{lr}
        R \frac{x_i}{\sum_{j=1}^{n}x_j}, & \text{if } \max(x_1,\ldots,x_n)>0 \\
        0, & \text{if }  \max(x_1,\ldots,x_n)=0 
        \end{array}
        \right.
\end{equation}

This removes the $\xeq=0$ equilibrium without significantly affecting other equilibria. 
In the two-player case, an equilibrium still involves the two players contributing the same value $x$ subject to different constraints and under the additional assumptions that $R<\min(v_1,v_2)$.
This expresses the fact that the player's valuations of the system must be at least greater than the value of the reward -- it would make little sense to gain a reward that is greater than the value of the system functioning.
We prove the following theorem in Appendix~\ref{proof:reward-ne}.

\begin{theorem}\label{thm:reward-ne}
In a two player NTEG with reward $R<\min(v_1,v_2)$ there exist infinite Nash equilibria where both players contribute the same value $x$ such that

\begin{equation}
    \left\{
                \begin{array}{ll}
                 
                 \frac{c_1}{4R}((R\frac{1-\sqrt{\Delta'_1}}{2c_1})^2-\beta_1^2)<
                 x<\frac{c_1}{4R}((R\frac{1+\sqrt{\Delta'_1}}{2c_1})^2-\beta_1^2)\\
                 \frac{c_2}{4R}((R\frac{1-\sqrt{\Delta'_2}}{2c_2})^2-\beta_2^2)<
                 x<\frac{c_2}{4R}((R\frac{1+\sqrt{\Delta'_2}}{2c_2})^2-\beta_2^2)                 
                \end{array}
              \right.
  \end{equation}
with $\Delta'_1=1+4\frac{c_1}{R}(\frac{v_1^2c_1}{R}+\frac{v_1}{c_1})$ and $\Delta'_2=1+4\frac{c_2}{R}(\frac{v_2^2c_2}{R}+\frac{v_2}{c_2})$.
\end{theorem}

We leave the multiplayer analysis as future work.

\subsection{Social optimum}
An insight from Varian's work is that the equilibria and social optima are not necessarily the same e.g., the total effort social optimum involves players contributing much more than in the Nash equilibrium~\cite{varian2004system}.

In the NTE case, the social optimum is for players to contribute the smallest non-zero amount possible as this maximises the level of decentralisation while minimising their costs.
If all contributions are equal then in most cases it is also a Nash equilibrium.
This convenient outcome is expected from our choice of NTE that reflects a desire to ensure that the social goal of decentralisation is met, so the NTE function is well defined in that sense.
The only exception is when the benefit-cost ratio of some players is too low as they then free-ride.

Figuring out an acceptable minimal contribution can be straightforward when it is possible to impose a minimum contribution.
Ethereum's implementation of proof-of-stake does this, but not all systems impose a minimum contribution. 

\section{Robustness Of Decentralisation Conscious Players to Variations By Others}\label{sec:robustness}
In practice, players may leave or join the game, as well as increase or decrease their contributions because of selfish behaviour or, more generally, Byzantine faults.
Thus, it is important to analyse how decentralisation conscious players tolerate variations in the actions of other players.
We do this by studying how the equilibria for the NTEG change after such events.

In the analysis that follows we will be using a result derived in the proof of Theorem~\ref{thm:nash-equilibria}, which is that for each player $j$ the best response to (fixed) contributions of other players is as follows.

\begin{enumerate}
\item if $\sum_{i\ne j}x_i<\beta_j$, contribute $\min(\max_{i\ne j}(x_i),\beta_j-\sum_{i\ne j} x_i)$
\item if $\sum_{i\ne j}x_i\ge\beta_j$, contribute zero.
\end{enumerate}

Equivalently, player's $j$ best response can be written as in Equation~\ref{eq:multiplayer-bestresponse}.
\begin{equation}\label{eq:multiplayer-bestresponse}
\max\{ 0,\min(\max_{i\ne j}(x_i),\beta_j-\sum_{i\ne j} x_i)\}
\end{equation}

In this section, we note $n$ the number of contributing players.
\subsection{Change in number of players}

\subsubsection{New player joining} 
\paragraph*{1-value equilibrium} 
We first consider the case where the players play the 1-value equilibrium described in Theorem~\ref{thm:nash-equilibria}.
If one player joins the game, the robustness of the equilibrium depends both on $\xeq$ and on the benefit-cost ratio $\beta$ of the new player.
From Condition~\ref{eq:multiplayerNE}, we have that $\xeq\le\frac{1}{n}\beta_j,\ \forall 1\le j\le n$.

We proceed as follows.
For different values of $\xeq$ we study what would be the new player's best response $x_{\new}$ and whether they would join the game i.e., contribute a non-zero effort.
We then look at whether the introduction of a new player playing $x_{\new}$ disrupts the equilibrium for the rest of the players i.e., whether having $n$ players play $\xeq$ and one player play $x_{\new}$ is still an equilibrium.
We find that the original $n$ players change their contributions if and only if $\beta>\beta_1$ and $\frac{1}{n+1}\beta_1<\xeq<\frac{1}{n}\beta$.
We prove this result in Appendix~\ref{proof:robustness-new-player}.

\begin{theorem}\label{thm:robustness-new-player}
In a NTEG that is in a state of 1-value equilibrium with $n$ players contributing $\xeq$, the introduction of a new player with benefit-cost ratio $\beta$ changes the value played by the other players if and only if $\beta>\beta_1$ and $\frac{1}{n+1}\beta_1<\xeq<\frac{1}{n}\beta$.
\end{theorem}

\paragraph*{2-value equilibrium}
In the case where the players were initially in a 2-value equilibrium, we have the following theorem, which we prove in Appendix~\ref{proof:newplayer2value}.

\begin{theorem}\label{thm:newplayer2value}
In a NTEG that is in a state of 2-value equilibrium with player 1 playing $x_1$ and the other $n-1$ players playing $\xeq$, the introduction of a new player with benefit-cost ratio $\beta$ does not change the value played by the other players unless $\sum_{i=1}^n x_i <\beta$ or $\beta_1<\beta$.
\end{theorem}

We now consider how the utility of each player changes following the introduction of a new player.
If the new player contributes a strictly positive effort and the value played at equilibrium stays unchanged for the other players it is clear that the introduction of a new player increases everyone's utility as it increases the reliability of the system without changing anyone's cost.
When the equilibrium is changed, if only one player (player 1) leaves the system, then this is simply a player replacement and the reliability of the system stays the same.
Player 1 increases their utility in this case as the reliability of the system is the same as before but their cost is now zero.

However, from the proof of Theorem~\ref{thm:robustness-new-player} we have that a new player could potentially incentivise more than one player to decrease their contribution.
Lemma~\ref{cor:maxtwocontrib} tells us that this means that the players would potentially need many iterations before reaching a new equilibrium if they reach one, where only one or two values are played.
Although it could be presumed that a new player joining should increase the reliability of the system, this result shows that if one or more players have to decrease their contributions then it is not clear that the final reliability of the system will be higher with $n+1$ player than with the original $n$ players.
We study simulations of equilibrium disruption in Section~\ref{sec:simulations} and leave a rigorous study of the outcome of the new game as an open problem.

\subsubsection{Player leaving the game} 
In the case where a player leaves the game, we have the following theorems, which we prove in Appendix~\ref{proof:player-leaves} and~\ref{proof:player-leaves-2}.

\begin{theorem}\label{thm:player-leaves}
In the NTEG, if the $n$ players are playing a $1-$value Nash equilibrium, the removal of a new player with benefit-cost ratio $\beta_i$ does not change the value played by the other players.
\end{theorem}

\begin{theorem}\label{thm:player-leaves-2}
In the NTEG with $n$ players playing a 2-value equilibrium where players 2 to $n$ play the same value $\xeq$ at the equilibrium, the removal of a new player with benefit-cost ratio $\beta_i$ changes the value played by the other players unless in the specific case where player $1$ is leaving the game.

\end{theorem}

If other contributions stay unchanged, a player leaving the system decreases the reliability of the system as it renders it more centralised.
The utilities of the remaining players will therefore always decrease in this case.

\subsection{Deviation from an equilibrium}\label{sec:deviation}

We now consider the case where one player (player $k$) deviates from the equilibrium and changes their contribution to $x_{k_0}$. 
We are concerned with the response of the $n-1$ other players and what new equilibrium is reached, regardless of whether it will be the best strategy for player $k$ to keep their value $x_{k_0}$ in the new equilibrium (i.e., player $k$ may be irrational).
We prove the following theorem in Appendix~\ref{proof:deviation1}

\begin{theorem}\label{thm:deviation1}
In the NTEG with $n$ players contributing the same value $\xeq$ at the equilibrium, the deviation of player $k$ with benefit-cost ratio $\beta_k$ to a new value $x_{k_0}$ does not change the value played by the other players unless $x_{k_0}>\xeq$.
\end{theorem}

In the 2-value equilibrium, the results are very similar.
We prove the following theorem in Appendix~\ref{proof:deviation2}.

\begin{theorem}\label{thm:deviation2}
In the NTEG with $n$ players playing a 2-value equilibrium where players 2 to $n$ play the same value $\xeq$ at the equilibrium, the deviation of player $k$ with benefit-cost ratio $\beta_k$ to a new value $x_{k_0}$ changes the value played by the other players unless in the specific case where player $1$ is deviating to a new value $x_{k_0}\ne x_1$ and for all $2\le j\le n$ we have
(1) $ x_{k_0}<\xeq$
(2) $\beta_j>(n-2)\xeq+x_{k_0}+\max(\xeq,x_{k_0})$ and
(3) $(n-2)x_{k_0}+x_{k_0} <\beta_2$.

\end{theorem}

In the case where the players do not change their equilibrium after an irrational player deviates (i.e., $x_{k_0}<x_{eq}$) the utility of players will decrease as reliability will be lower for the same costs and contributions.

In the other case, before the other players can adjust their contribution, their utility will also decrease, and in some realistic cases, players may not be able to change their contribution as we discuss in Section~\ref{sec:discussion}.
This is an undesirable effect defined as \emph{immunity} by Abraham et al.~\cite{abraham2006distributed} in the context of distributed systems where one or more irrational players can negatively impact the utility of rational players.
If players can change their contributions, the reliability functions could go up or down depending on the new value $x_{eq}$ and the benefit-cost of other players (i.e., whether they will free ride).

After the deviation from player $k$, we have from Condition~\ref{eq:multiplayer-bestresponse} that each player $i$ such that $x_{k_0} \ge \frac{\beta_i - (n-2)\xeq}{2}$ changes their contribution to $x_{i,\new}=\beta_i - x_{k_0} - (n-2)\xeq$ or zero if that value is negative, and each player $i$ such that $x_{k_0} \le \frac{\beta_i - (n-2)\xeq}{2}$ changes their contribution to $x_{k_0}$.

In Lemma~\ref{cor:maxtwocontrib}, we showed that if there exists two contributing rational players whose contributions are strictly less than $\max_{j}(x_j)$, then those players must have the same benefit-cost ratio.
This is true regardless of the existence of an irrational player.
Since we assume that all the benefit-cost ratios are different, this means that there can be at most one rational player playing strictly less than the maximum value $x_M$.
According to the strategy defined in Condition~\ref{eq:multiplayer-bestresponse}, no rational player is incentivised to play more than $\max_{j}(x_j)$.
Thus, even after players adjust their contributions we will still have $\max_{j}(x_j) = x_{k_0}$ and, following the deviation, the bulk of the players will align with the deviating players or free ride, except for one rational player.
By setting $x_{k_0}$ high enough, the deviating player could ensure that many players switch to free-riding, which could pose a threat to the system if it facilitates one party taking control of the system (e.g., a 51\% attack).

\subsection{Non-myopic players}
Motivated by Br\"unjes et al.~\cite{brunjes2018reward}, we consider \emph{non-myopic players} deviating from the equilibrium.
The utility function of such players accounts for the effects an action will have on the other players, unlike Nash equilibria that consider the best response of players given that the other players' strategies are fixed.

In the previous section, we have seen that a player deviating from the equilibrium may disrupt the best response of the other players and lead to a new equilibrium.
In a Nash equilibrium, assuming that other players keep their contribution unchanged, deviating means that one's utility is reduced, but this does not account for the possibility of a new equilibrium being reached.
A new equilibrium (if reached) may be a better equilibrium for the deviating player if their utility is higher in the new equilibrium.

Does a non-myopic player have incentives to deviate from the equilibria we have derived?
We have established that a condition to disrupt the equilibrium is to change one's contribution to a value $x_{k_0} > \xeq$.
We have also observed that by setting this value high enough, the deviating player can cause some players to free ride.
In a NTEG without a reward, the new equilibrium would, therefore, have fewer contributing players with greater contributions.
In a 1-value equilibrium, this would mean we have $F(x_1,\ldots,x_n)=n_{\text{\new}}<n$ where $n_{\text{\new}}$ is the new number of contributing players.
However, because $x_{k_0}>\xeq$, the cost will be higher and this strategy is therefore not rational as the new equilibrium results in less utility for the deviating player and the other players.

In a NTEG with reward, however, fewer players implies a greater share of rewards.
Thus a non-myopic player may be incentivised to deviate from an existing equilibrium to reach a new one with fewer contributing players.

This suggests that a fixed proportional reward may increase centralisation.
Designing a protocol with a variable reward such that players would earn similar revenue regardless of the number of players is an open problem due to the pseudonymous nature of systems like cryptocurrencies.
Another alternative is to rely on a fixed reward but design the system such that it is not possible to increase one's contribution, as in proof-of-personhood schemes~\cite{borge2017proof}.

\subsection{Coalition-resistance}
A group of miners may decide to form a coalition if this increases their expected gain, even if doing so centralises the system.
In this case a coalition is equivalent to having one player contributing $X=\sum_{i\in [i_1,\ldots,i_c]}x_i$ for all the players $(i_1,\ldots,i_c)$ in the coalition instead of having each contributing separately.
Because the sum of the efforts stay the same but the maximum effort potentially increases, $F(x_1,\ldots,x_{i_1},\ldots,x_{i_c},\ldots,x_n)\ge F(x_1,\ldots,X,\ldots,x_n)$.
Thus, the utility of decentralisation conscious players decreases when they form a coalition i.e., they are not incentivised to create coalitions.

\section{Dynamics Of Decentralisation Conscious Players}\label{sec:simulations}
As we have shown, there are many possible equilibria, each corresponding to different equilibrium values.
How an equilibrium is reached i.e., how quickly and how many players reach it, as well as which equilibrium value is reached could depend on several factors that we look at in this section.

\paragraph*{Methodology}
Using a Python script, we simulate the NTEG where each player computes their best strategy at each time unit. 
By iterating over multiple time units we observe how players (simultaneously) re-evaluate their contributions based on the effort of other players in the previous time unit.
The scenarios we simulate are not exhaustive but highlight interesting behaviour, the benefit-cost ratios were chosen randomly within a range.

\paragraph*{Random initial values}

To observe how an equilibrium is reached, we initialise a NTEG with 10 players to which we assign random initial values (contributions, costs, benefits) and look at how they change their contributions until an equilibrium is reached. (The same initial contributions and benefit-cost ratios are used for every simulation.)
According to the strategy defined by Equation~\ref{eq:multiplayer-bestresponse}, no decentralisation conscious player is incentivised to contribute more than other players hence the player that has the maximum contribution in step 1 of the game (set by nature's move) will be reducing their contribution in the next step.
On the other hand, other players with a high enough benefit-cost ratio will be incentivised to increase their contributions to the maximum value in step two of the game.

\begin{figure}[t]
\begin{subfigure}{.321\textwidth}
\centerline{\includegraphics[width=\linewidth]{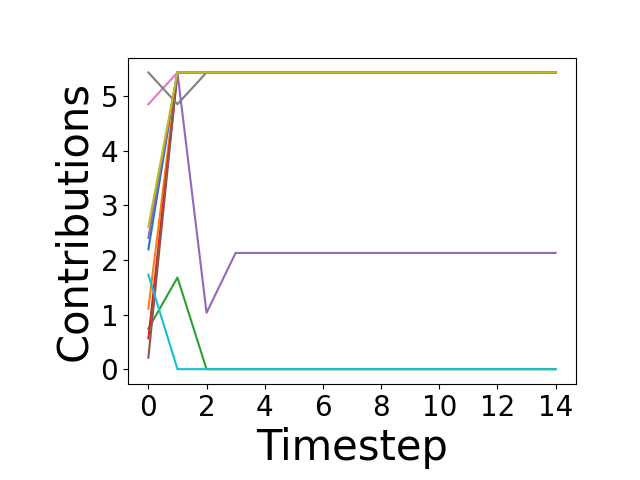}}
\caption{}
\label{fig:two-value-eq}
\end{subfigure}
\begin{subfigure}{.329\textwidth}
\centerline{\includegraphics[width=\linewidth]{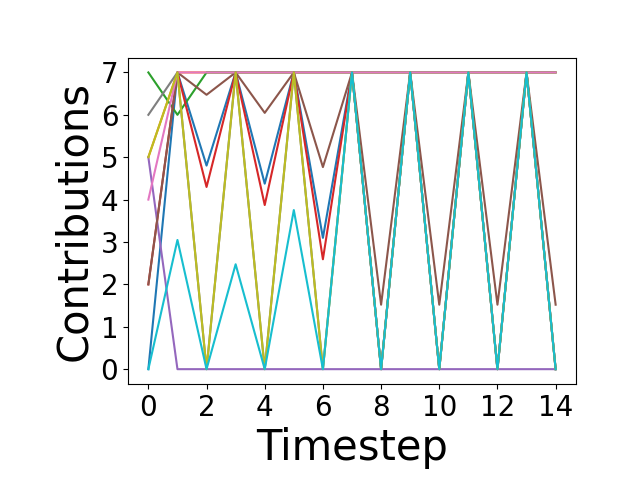}}
\caption{}
\label{fig:eqnotreached}
\end{subfigure}
\caption{Without constraints on contribution changes players can reach an equilibrium (Figure~\ref{fig:two-value-eq}) but may also oscillate indefinitely (Figure~\ref{fig:eqnotreached}).}
\end{figure}

Under ideal conditions i.e., when the maximum contribution $x_{\max}$ is such that $nx_{\max}<\min_i\beta_i$, the equilibrium is reached after a few steps.
Players with the greatest benefit-cost ratios align their contributions to the maximum value (except perhaps for one of them, resulting in a 2-value equilibrium) while the remaining players free ride, as shown in Figure~\ref{fig:two-value-eq}.

In other cases, as shown in Figure~\ref{fig:eqnotreached}, some players may keep oscillating indefinitely.
For these players, it must be the case that $\beta_j<n \xeq$, else playing $\xeq$ at the same time as other players will be their best strategy and an equilibrium will be reached.
Thus whenever everyone is playing $\xeq$ at one time unit, they decrease their contributions to $\beta_j - n \xeq$ in the next step.
However, after other oscillating players have also decreased their contributions, it is now the best strategy to go back to $\xeq$, and so on.
This is due to players being myopic, not anticipating that other players will increase their contributions at the same time as them.

\paragraph*{Constraints on the rate of change of contributions.}
To avoid the unrealistic case where players oscillate forever we constrain the change in each player's contribution from one time unit to another by a factor $\Delta$. 
This dampens the oscillations and allows players to converge to an equilibrium.

\begin{figure}[t]
\begin{subfigure}{.329\textwidth}
\centerline{\includegraphics[width=\linewidth]{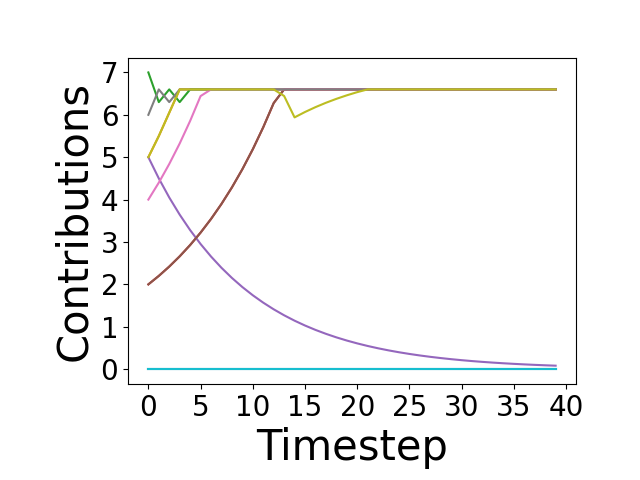}}
\caption{$\Delta = 0.1$}
\label{fig:cons0.1}
\end{subfigure}
\begin{subfigure}{.329\textwidth}
\centerline{\includegraphics[width=\linewidth]{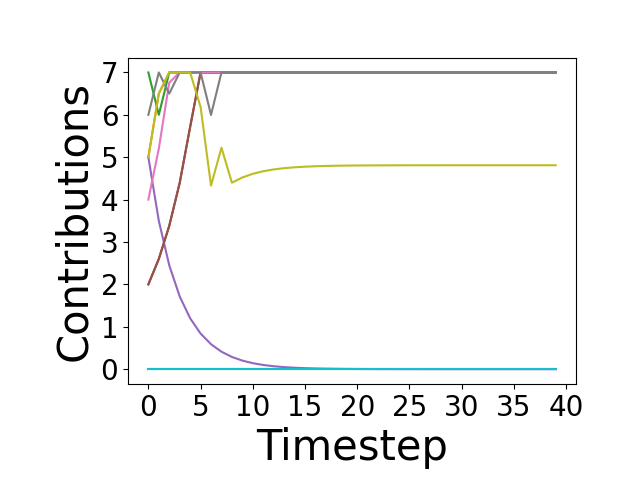}}
\caption{$\Delta = 0.3$}
\label{fig:cons0.3}
\end{subfigure}
\begin{subfigure}{.329\textwidth}
\centerline{\includegraphics[width=\linewidth]{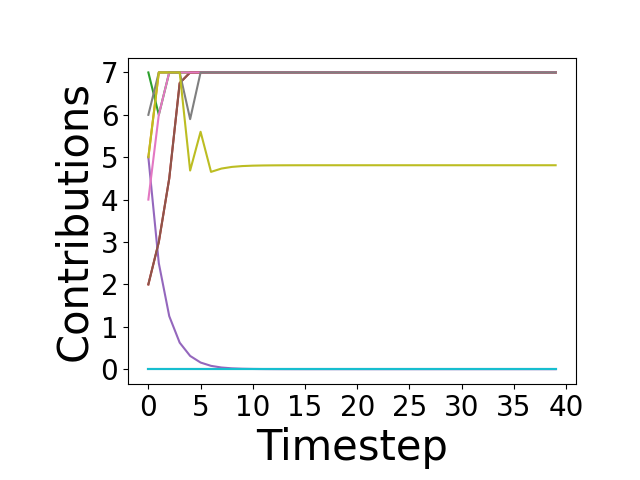}}
\caption{$\Delta = 0.5$}
\label{fig:cons0.5}
\end{subfigure}
\caption{Oscillations disappear with constraints on contribution changes. The speed at which equilibria are reached depends on the constraints (slower with $\Delta = 0.1$, faster with $\Delta = 0.5$), as do the type of equilibria (1-value with $\Delta = 0.1$, 2-value $\Delta = 0.3$ or $0.5$) and equilibrium value (greater with $\Delta = 0.3$ or $0.5$). }
\end{figure}

Because $\Delta$ affects how quickly players can converge to an equilibrium, the equilibrium that is reached varies with $\Delta$.
For example, in the case where $\Delta=0.1$ participants are allowed to change their contributions by at most 10\% from one time unit to another and a $1$-value equilibrium is reached, as shown in Figure~\ref{fig:cons0.1}.
When $\Delta = 0.3$ or $\Delta = 0.5$, a $2$-value equilibrium is reached, as shown in Figures~\ref{fig:cons0.3} and~\ref{fig:cons0.5}.
Keeping this in mind we will, however, stick to the $\Delta=0.1$ case in most of the simulations that follow for simplicity as the overall player behaviours i.e., players increasing their contribution or free-riding are the same although the final equilibrium differs.

We have also computed the different values of the reliability in each case but did not observe any clear pattern.
Whether there is a pattern that is not clearly observable is left as an open problem.

Not only is a constraint on the change in the effort of players useful for them to efficiently converge to an equilibrium, it is also realistic. 
Players in real life are likely to understand the adverse effects of over correcting and are also likely to have constraints on how much they can change their effort (at least upwards) due to the cost of doing so.
We discuss this constraint further in the next section, in relation to resource scarcity.

Moreover, every player updating their contributions at the same time is not a realistic assumption either.
Bounding the change of contribution of each player from one step to another also helps get closer to a continuous time model.

\begin{figure}[t]
\begin{subfigure}{.329\textwidth}
\centerline{\includegraphics[width=\linewidth]{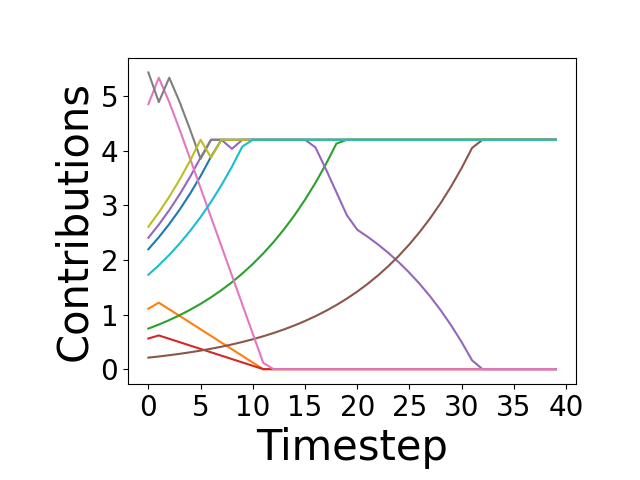}}
\caption{$\Delta = 0.1$}
\label{fig:cons0.1tot}
\end{subfigure}
\begin{subfigure}{.329\textwidth}
\centerline{\includegraphics[width=\linewidth]{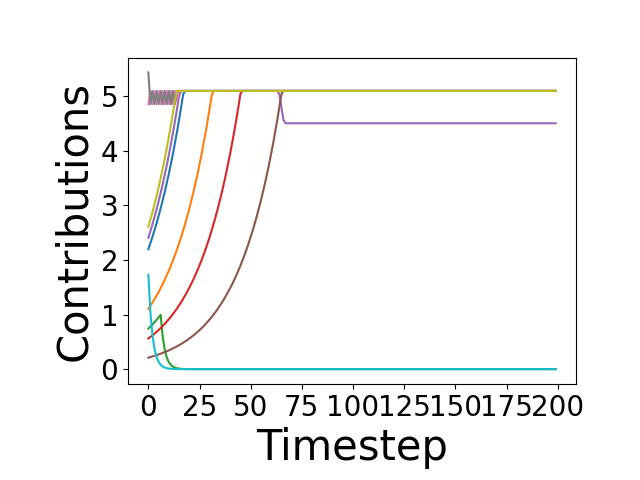}}
\caption{$\Delta_+ = 0.05$, $\Delta_- = 0.4$}
\label{fig:up-down-1}
\end{subfigure}
\begin{subfigure}{.329\textwidth}
\centerline{\includegraphics[width=\linewidth]{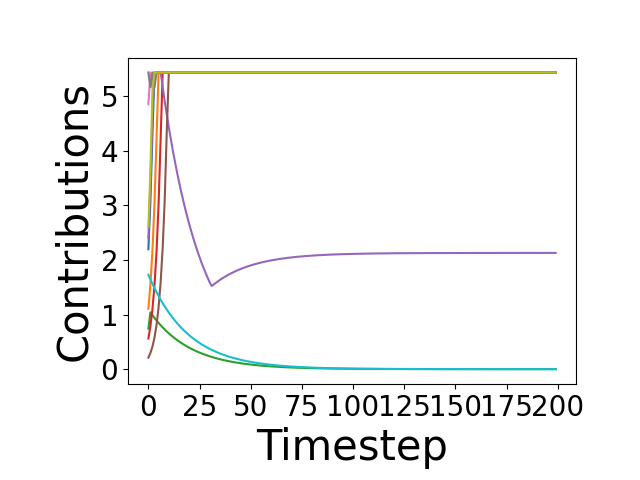}}
\caption{$\Delta_+ = 0.4$, $\Delta_- = 0.05$}
\label{fig:up-down-2}
\end{subfigure}
\caption{Constraining the total effort can increase free-riding and reduce equilibrium values (Figure~\ref{fig:cons0.1}), as can reductions in contributions being easier than increases (Figures~\ref{fig:up-down-1} and~\ref{fig:up-down-2}).}
\end{figure}

\paragraph*{Constraints on total effort}
Another constraint that can be implemented is a limit on the overall change in the effort of all players i.e., the total effort.
This models the constraint that the stock of resources used to contribute effort (e.g., new hardware) may be limited at any point in time.
Figure~\ref{fig:cons0.1tot} shows that in this case, some players may not be able to change their contribution enough to converge to the equilibrium and, therefore, switch to free riding.

Since it is usually easier to reduce one's contribution than to increase it, we also simulate the game with different constraints on the increase and the decrease of contributions from one step to another. 
We see in Figures~\ref{fig:up-down-1} and~\ref{fig:up-down-2} that a $2$-value equilibrium is reached, although the relative constraints on increasing and decreasing contributions result in different equilibrium values.
The value played by the bulk of the player $\xeq$ is higher when there is a greater constraint on the increase of contributions than on the decrease.
This is because players can more rapidly reach the new maximum value.
As a consequence, the second value played at the equilibrium is smaller.

\begin{figure}[t]
\begin{subfigure}{.329\textwidth}
\centerline{\includegraphics[width=\linewidth]{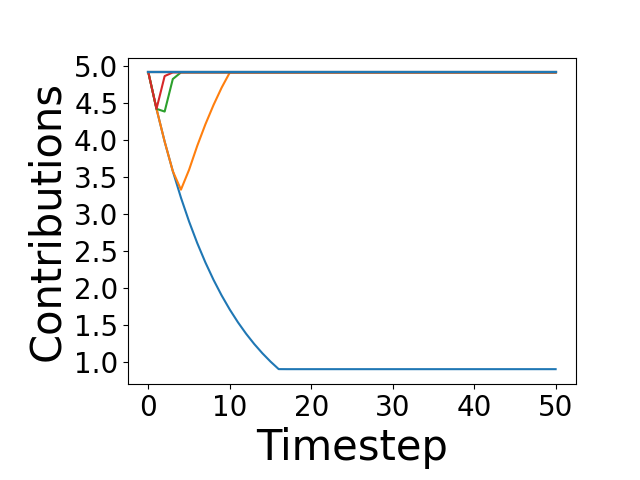}}
\caption{$\Delta = 0.1$, a new player joins}
\label{fig:newplayer-constraint}
\end{subfigure}
\begin{subfigure}{.329\textwidth}
\centerline{\includegraphics[width=\linewidth]{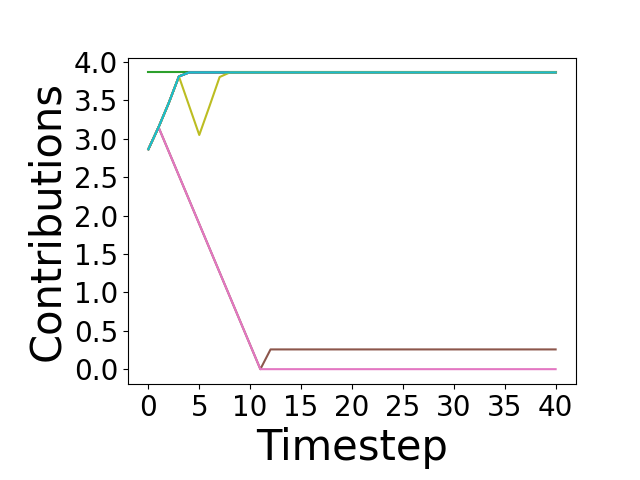}}
\caption{$\Delta = 0.1$, a player deviates}
\label{fig:playerdeviating-constraint}
\end{subfigure}
\caption{Disruptions to an equilibrium due to a new player joining or a player deviating lead to new equilibriums.}
\end{figure}

\paragraph*{Disruptions to an equilibrium}
A new player joining the game when it is in a 1-value equilibrium (which happens according to the conditions defined in Theorem~\ref{thm:robustness-new-player}) can lead to a new equilibrium being reached after a few steps, as shown in Figure~\ref{fig:newplayer-constraint} in the case of a strong constraint.

When an equilibrium is disrupted by a player deviating from the equilibrium, players that increase their contribution to contribute more effort than the equilibrium value incentivise other decentralisation conscious players to free ride or increase their effort to reach a new equilibrium value if it is allowed by their benefit-cost ratio.
This is shown in Figure~\ref{fig:playerdeviating-constraint}, in the case of a strong constraint.

\section{Discussion}\label{sec:discussion}
\subsection{The role of decentralisation conscious players}
Our model and choice of NTE function shows that decentralisation conscious players can help maintain a decentralised system.
However, as Theorems~\ref{thm:deviation1} and~\ref{thm:deviation2} show, decentralisation conscious players only ever increase their effort in response to another player increasing their contribution at the cost of decentralisation.
They maintain decentralisation within the constraints of their benefit-cost ratio but ignore players that free ride after their benefit-cost ratio no longer allows them to contribute.

Because decentralisation conscious players can only maintain a pre-existing level of decentralisation and can be leveraged by selfish players to implement a minimum benefit-cost ratio that acts as a form of gate-keeping against players with lower benefit-cost ratios, there is a distinction between decentralisation conscious players and \textit{altruistic} players that operate regardless of their benefit-cost ratio.
This suggests that new mechanisms dictating how effort is contributed or rewarded may be needed for players to have rational ways of increasing decentralisation outside of purely altruistic behaviour.

\subsection{Modelling constraints}
\paragraph*{Resource scarcity}
Players contribute based on their benefit-cost ratios and, as we have seen in Section~\ref{sec:simulations}, equilibria depend on the rate of change of contributions.
An implicit assumption made by our model is that a player can contribute more (at a cost) should they wish to do so but this may not be possible.
For example, cryptocurrency mining hardware has suffered from shortages that forced buyers to obtain hardware at significant premiums and logistical difficulties~\cite{boeing}.
When resources are unobtainable, it can become impossible to contribute more or continue contributing the same amount (if resources must be replaced), causing involuntary deviations from otherwise rational strategies.

If it is impossible to acquire the resources to contribute, the system will rely on players having a high valuation of the system.
Contributors to systems like Tor~\cite{syverson2004tor} operating nodes at a loss may demonstrate this but in the case of cryptocurrencies new miners are less likely to have a high valuation of the system because they are unlikely to have a stake in it, unlike miners that have accumulated rewards.
Miners in cryptocurrencies that are more centralised due to the high practical costs of mining can, therefore, form an effective oligopoly~\cite{gencer2018decentralization,arnosti2018bitcoin}.

Can we avoid issues of resource scarcity?
One way of avoiding the problematic reliance on resources with variable stock (e.g., stake, hardware) is to opt for mechanisms like proof-of-personhood~\cite{borge2017proof}, which is equally distributed (``1 person = 1 vote'') and maximises the NTE, although this has other issues to overcome.

\paragraph*{Geographical and political decentralisation }
Because players contribute based on their benefit-cost ratios so the geographical distribution of players will matter if costs vary with location.
For example, cryptocurrency mining is concentrated in the few areas where mining is most profitable.

Markets are also affected by political power and changes in regulations.
China controlled 65\% of Bitcoin's hashpower in 2019~\cite{coindesk_2} but following new Chinese regulations~\cite{coindesk_3} the share of hashpower in the US has grown due to political stability with respect to Bitcoin mining~\cite{coindesk_1}.
The impact of markets and political power on decentralisation adds complexity and uncertainty in models, which may motivate decentralised systems less reliant on other markets e.g., proof-of-stake (based on the cryptocurrency's native tokens) or proof-of-personhood may be easier to reason about than proof-of-work (energy and hardware markets).

\paragraph*{Incomplete and unequal information}
Our model has assumed perfect information at each step with players changing their contributions based on this information, but players could hide information such as the stock of unused resources they have at their disposal.
Attacks such as selfish mining in proof-of-work cryptocurrencies~\cite{eyal2014majority} are based on abusing information asymmetry, as are hostile takeovers which use previously unused but available mining capacity~\cite{bonneau2018hostile}.
There is also an inherent delay in information propagating through a network.
This may result in different equilibria as players adapt their contributions based on the information they receive at a point where it may no longer be accurate.

How much this matters is hard to determine.
Attacks such as selfish mining have seldom been observed, and effort rarely varies across short time periods. (See the Bitcoin hashrate distribution over short time periods, even if the larger trend is growth~\cite{blockchaincom}.)
This may be due to issues like acquiring the additional resources needed to contribute more effort, but it may also be to maintain a level of decentralisation as our model suggests miners might do. 

\subsection{Related Work}
\label{sec:related}

There is an important literature on modelling incentives in cryptocurrencies through refinements of Nash Equilibria that has been systematised~\cite{azouvi2019sok}.
Although the types of players and games considered vary across papers, none of the papers surveyed (except Varian's paper~\cite{varian2004system}) consider the reliability of the system.

Varian's system reliability paper~\cite{varian2004system} has previously been extended by Grossklags et al.~\cite{grossklags2008secure} in the context of investments in security and insurance.
Grossklags et al.~\cite{grossklags2010information} have also applied Varian's model to study the difference between expert and naive players in security games to quantify the impact of information.
In this work, we have instead focused on decentralisation and introduced the NTEG, which extends Varian's model in another direction.

\section{Conclusion}
\label{sec:conclusion}
We have proposed a model for decentralisation conscious players based on the NTE function we have introduced.
The Nash equilibria show what could be expected from such players.
Using simulations we have also considered how players may reach an equilibrium, including after disruptions.
There is a variety of possibilities for future work and opportunities to apply our model to specific cases.
This includes cases with valuations of the system which are hard to precisely define e.g., ideological commitment, as well as cases with very explicit valuations and dependencies on rewards but complex financial optimisation such as cryptocurrencies.
Protocol designers who wish to incorporate rational players, as opposed to honest players, but also wish to incorporate the reliability of the system in addition to short-term rewards could use the NTE function.

\section*{Acknowledgments}
Alexander Hicks was partially supported by Protocol Labs for this work.

\bibliographystyle{splncs04}
\bibliography{refs}
\appendix
\section{Proof of Theorem~\ref{thm:two-player-ne}} \label{proof:two-player-ne}
To determine player 1's best strategy, we first study the continuous function $x_1\mapsto u_1(x_1)$ for a fixed $x_2$ and try to find its global maximum.
There are two cases to consider, depending on whether $x_1 \leq x_2$ or $x_1>x_2$.

If $x_1< x_2$, we have $u_1 = \ln(\frac{x_1+ x_2}{x_2})v_1- c_1x_1$ and $\frac{du_1}{dx_1}=\frac{v_1}{x_1+x_2}-c_1$, so $\frac{du_1}{dx_1} \ge 0$ iff Condition~\ref{firstcasecond} holds.

\begin{equation}\label{firstcasecond} x_1+x_2\le\beta_1 \end{equation}

If $x_1>x_2$, we have $u_1 = \ln(\frac{x_1+ x_2}{x_1})v_1- c_1x_1$ and $\frac{du_1}{dx_1}=v_1(\frac{1}{x_1+x_2}-\frac{1}{x_1})-c_1$, so $\frac{du_1}{dx_1}<0$.

We can, therefore, define player 1's strategy as follows.
\begin{enumerate}
  \item If $\beta_1 < x_2$ then (from condition~\ref{firstcasecond}) $u_1$ is a decreasing function and player 1's best strategy is to play $x_1 =0$.
  \item If $x_2\le\beta_1$ then (from condition~\ref{firstcasecond}) $u_1$ is increasing up to $\min(x_2,\beta_1-x_2)$ so there are two cases to consider.
  \begin{enumerate}
    \item If $x_2\le \beta_1-x_2$ i.e., if $x_2\le \frac{1}{2}\beta_1$, then condition~(\ref{firstcasecond}) is always satisfied for $x_1\le x_2$ and hence player 1 best strategy is to play $x_1 =x_2$. When $x_1>x_2$ $u_1$ decreases so player 1 maximises $u_1$ by playing $x_2$.
    \item If $x_2>\beta_1-x_2$ or equivalently if $\frac{1}{2}\beta_1< x_2\le \beta_1$: then player 1 best strategy is to play $x_1 = \beta_1-x_2$ (after which $u_1$ starts decreasing according to condition~(\ref{firstcasecond})).
  \end{enumerate}
\end{enumerate}

The same analysis can be repeated for $u_2$, giving the same results.
Therefore, there exist infinite equilibria where $x_1=x_2=\xeq$ and $\xeq \le \frac{1}{2}\min(\beta_1,\beta_2)$.
We now show that these are the only equilibria of the game.

If one of the contributions, say $x_2$, is zero, then we are in case (2a) of the strategy for player 1 and their best response is also zero.
Thus, we have $x_1 = x_2 =\xeq$ such that $x\le \frac{1}{2}\min(\beta_1,\beta_2)$.

For the rest of this proof, we assume that $x_1,x_2>0$.
Proceeding by contradiction, we assume that there exists a Nash equilibrium $(x_{1_0},x_{2_0})$ for which $x_{2_0}\ne x_{1_0}$.
Since $x_{2_0}\ne x_{1_0}$ and $x_{1_0}>0$ by assumption, this means we are in case (2b) of player 1's strategy and player 1 maximises $u_1$ when $x_1=\beta_1-x_{2_0}=x_{1_0}$.

Similarly, we have $\beta_2-x_{1_0}=x_{2_0}$.
Solving this system of two equations gives us $x_{1_0}+x_{2_0}=\beta_1=\beta_2$.
So unless $\beta_1=\beta_2$, we have a contradiction.

If $\beta_1=\beta_2=v/c$, then we have $x_{1_0}+x_{2_0}=v/c$.
Again, proceeding by contradiction, we assume that $x_{1_0}>\frac{1}{2}v/c$.
This implies that $x_{2_0}<\frac{1}{2}v/c$, implying that we are in case (2a) of player's 1 strategy and thus that player 1's best response is $x_{1_0}=x_{2_0}$.
The analysis is the same if $x_{1_0}<\frac{1}{2}v/c$, applied to player 2.
This proves that players contribute the same at the equilibrium.

\section{Proof of Theorem~\ref{thm:nash-equilibria}}\label{proof:nash-equilibria}

The analysis of $x_1\mapsto u_1(x_1)$ for $(x_2,\ldots,x_n)$ fixed is similar to the analysis done in the two player case.

If $x_1<\max_{i=2}^n(x_i)$, we have $u_1 = \ln(\frac{x_1+ \sum_{i=2}^nx_i}{m})v_1- c_1x_1$ and $\frac{du_1}{dx_1}=\frac{v_1}{\sum_{i=1}^n{x_i}}-c_1$.
Hence, $\frac{du_1}{dx_1}\ge0$ iff Condition~\ref{multiplayerfirstcase} holds.

\begin{equation} \label{multiplayerfirstcase} \sum_{i=1}^n x_i\le\beta_1 \end{equation}

If $x_1>\max_{i=2}^n(x_i)$, we have $u_1 = \ln(\frac{x_1+  \sum_{i=2}^n x_i}{x_1})v_1- c_1x_1$ $\frac{du_1}{dx_1}=v_1(\frac{1}{\sum_{i=1}^n x_i}-\frac{1}{x_1})-c_1$, so $\frac{du_1}{dx_1}<0$.

As in the two player case, we can establish that player 1's best strategy for a fixed $(x_2,\ldots,x_n)$ is as follows.
\begin{enumerate}
\item If $\max_{i=2}^n(x_i) + \sum_{i=2}^n x_i\le\beta_1$, contribute $\max_{i=2}^n(x_i)$. This is because we will always have $\sum_{i=1}^n x_i\le \beta_1$ as long as $x_1\le\max_{i=2}^n(x_i)$, thus $u_1$ increases as $x_1$ increases up to $\max_{i=2}^n(x_i)$ and then decreases. The best response is thus $x_1=\max_{i=2}^n(x_i)$.
\item If $\sum_{i=2}^n x_i<\beta_1<\max_{i=2}^n(x_i)+\sum_{i=2}^n x_i$, contribute $\beta_1-\sum_{i=2}^n x_i$. This is because (from Condition~\ref{multiplayerfirstcase}) $u_1$ is increasing up to that point, then decreasing.
\item If $ \sum_{i=2}^n x_i\ge\beta_1$ then $u_1$ is decreasing and thus player 1's best response is to contribute nothing: $x_1 = 0$.
\end{enumerate}

The best strategy of every other player is derived in the same way.

Due to Condition~\ref{multiplayerfirstcase}, we have that for every contributing player, the following holds at the equilibrium.
\begin{equation}\label{eq:app-eq-cond}
\forall i\in [1,n]:\; \sum_{j=1}^n x_j \le \beta_i
\end{equation}

Additionally, from the strategy defined above, we have that the case where players $i+1$ to $n$ (for any $1\le i\le n-1$) contribute the same value $x$ such that $\frac{1}{n-i}{\beta_{i}}\le x\le\frac{1}{n-i}\beta_{i+1}$ and others free-ride is a Nash equilibrium.
All the players $i+1$ to $n$ are in case (1) of their strategies and thus contribute $x$ whereas all the players 1 to $i$ are in case (3) and contribute zero. 
It is straightforward that these are the only type of equilibria that exist where all contributing players contribute the same value.

Now, assume that there exist at least two contributing players $i_1$ and $i_2$ who contribute two values $x_{i_1}$ and $x_{i_2}$ at the equilibrium such that $x_{i_1} \ne x_{i_2}$.
Without loss of generality we assume $x_{i_1} < x_{i_2}$.
We start by showing that for every other contributing player $i$, their equilibrium contribution $x_i$ is equal to $x_{i_2}$. 
We show this by contradiction: we assume that $x_i\ne x_{i_2}$.

If $x_i<x_{i_2}$, then $x_i < \max_{j\ne i} x_j$ so we must be in case (2) of player $i$'s strategy defined above ($x_i>0$ by assumption).
We conclude that $x_i = \beta_i - \sum_{j\ne i} x_j$, which tells us that $\beta_i = \sum_{j=1}^n x_j$.
In a similar way, we have that $x_{i_1} = \beta_{i_1} - \sum_{j\ne i_1} x_j$ and thus $\beta_{i_1}=\sum_{j=1}^n x_j = \beta_i$, so players $i_1$ and $i$ have the same benefit-cost ratio
which contradicts our assumption (Definition~\ref{def:game}).
This also shows that there cannot exist two contributing players whose contributions are strictly less than $\max_j(x_j)$ unless they have the same benefit-cost ratio.
(This will later be used as Lemma~\ref{cor:maxtwocontrib}.)

It must be that $x_i>x_{i_2}$, then $x_{i_2} < \max_j x_j$ and the exact same analysis as above can be applied to $x_{i_2}$.
This leads us to $\beta_{i_2}= \beta_{i_1}$.
This is a contradiction, meaning that $x_{i}\le x_{i_2}$ and, therefore, $x_i=x_{i_2}$.

Thus at the equilibrium there can exist only two different non-zero contributions possible.

We assume that there exists $1$ players contributing $x_{i_1}$ and $M=n-1$ contributing $x_{i_2}$.
According to player $i_1$'s' strategy (case (2)) we have the following.
\begin{align}
x_{i_1} &= \beta_{i_1}  - M x_{i_2}\\
\end{align}

Because player ${i_1}$ is a contributing player, we have by assumption that $x_{i_1}>0$, which implies $x_{i_2}<\frac{\beta_{i_1}}{M}$.
We also have by assumption that $x_{i_1}<x_{i_2}$, which implies $\beta_{i_1}-Mx_{i_2}<x_{i_2}$, which in turn implies $\frac{\beta_{i_1}}{n}<x_{i_2}$.

According to condition~\ref{multiplayerfirstcase}, applied to player $i_2$, 
we have that $Mx_{i_2}+ x_{i_1}\le \beta_{i_2}$.
Replacing the value of $x_{i_1}$ in this inequality leads to $\beta_{i_1}\le \beta_{i_2}$.
We, therefore, have $\frac{\beta_{i_1}}{n} < x_{i_2}< \frac{\beta_{i_1}}{M}\le \frac{\beta_{i_2}}{M}$.
 
It is straightforward to verify that having one player with the smallest benefit-cost ratio play $x_{i_1}$ and the rest plays $x_M$ is a Nash equilibrium.

\section{Proof of Theorem~\ref{thm:reward-ne}}\label{proof:reward-ne}
With the addition of the reward, the utility function for player 1 is the following.
 \begin{equation}\label{eq:utility-reward}
 u_1(x_1)=
    \left\{
                \begin{array}{ll}
                  \ln(\frac{x_1+ x_2}{\max(x_1,x_2)})v_1- c_1x_1 + R \frac{x_1}{x_1+x_2}
 \text{ if } \max(x_1,x_2)>0 \\
                 0 \text{ if } \max(x_1,x_2)=0
                \end{array}
              \right.
  \end{equation}

If $x_1<x_2$, $u_1 = \ln(\frac{x_1+ x_2}{x_2})v_1- c_1x_1 + R \frac{x_1}{x_1+x_2}$ and $\frac{du_1}{dx_1}=\frac{v_1}{x_1+x_2}-c_1 + R \frac{x_2}{(x_1+x_2)^2}$.

We solve the inequality $\frac{du_1}{dx_1}>0$, written in terms of $X=x_1+x_2$ with $X>0$.
\begin{equation}\label{eq:rwd-condition}
\frac{du_1}{dx_1}>0 \Leftrightarrow X^2 -\frac{v_1}{c_1}X-\frac{Rx_2}{c_1}<0 
\end{equation}

Consider $\Delta_1 = (\frac{v_1}{c_1})^2+4\frac{Rx_2}{c_1}$ and $X_\pm =\frac{1}{2}(\frac{v_1}{c_1} \pm \sqrt{\Delta_1})$, where $X_\pm$ are the roots of the quadratic equation $X^2 -\frac{v_1}{c_1}X-\frac{Rx_2}{c_1}=0$.
We can rewrite inequality~\ref{eq:rwd-condition} in terms of $X_\pm$ as $(X-X_{-})(X-X_{+})<0$.
Since $X_{-}<X_{+}$, the solution to this inequality is $X_{-}<X<X_{+}$.
We also note that $\sqrt{\Delta_1}>\frac{v_1}{c_1}$ and hence $X_{-}<0$.
Since we also have $X\ge 0$, we can conclude the following for a fixed $x_2$.
\begin{equation}\label{eq:rwd-condition2}
\frac{du_1}{dx_1}>0 \Leftrightarrow  x_1 + x_2 < \frac{1}{2}(\frac{v_1}{c_1}+\sqrt{\Delta_1})
\end{equation}

If $x_1>x_2$, $u_1 = \ln(\frac{x_1+ x_2}{x_1})v_1- c_1x_1 + R \frac{x_1}{x_1+x_2}$ and $\frac{du_1}{dx_1}=v_1(\frac{1}{x_1+x_2}-\frac{1}{x_1}) -c_1 + R \frac{x_2}{(x_1+x_2)^2}$.
In that case we have $\frac{du_1}{dx_1}>0$ iff $\frac{x_2(x_1(R-v_1)-v_1x_2)}{x_1(x_1+x_2)^2}-c_1>0$ and, therefore, $\frac{du_1}{dx_1}<0$ if $R<v_1$.

\sarah{need to fix this}
Assume that $(x_{1_0},x_{2_0})$ are the equilibrium values.
If $x_{1_0}  > \frac{1}{2}(\frac{v_2}{c_2}+\sqrt{\Delta_2})= \frac{1}{2}(\frac{v_2}{c_2}+\sqrt{(\frac{v_2}{c_2})^2+4\frac{Rx_{1_0}}{c_2}})$ then $x_2\mapsto u_2$ is decreasing (due to the result derived above and the assumption that $R<\min(v_1,v_2)$) and thus player 2 maximises their utility by contributing $x_{2_0}=0$.
If $x_{2_0}=0$, player 1 is better off contributing a very small amount to be sure to get the reward while minimising their cost, so $x_{1_0}\approx 0$.
This contradicts the condition $x_{1_0} > \frac{1}{2}(\frac{v_2}{c_2}+\sqrt{\Delta_2})$, which means that $x_{1_0} < \frac{1}{2}(\frac{v_2}{c_2}+\sqrt{\Delta_2})$.
The exact same argument can be made to derive $x_{2_0} < \frac{1}{2}(\frac{v_1}{c_1}+\sqrt{\Delta_1})$.

Assume now that $x_{1_0}\ne x_{2_0}$ and, without loss of generality, that $\max(x_{1_0},x_{2_0})=x_{1_0}$.
Since $R<v_1$ then $\frac{du_1}{dx_1}<0$ for $x_1\ge x_{2_0}$ and hence player 1 best strategy is to contribute $x_{1_0}\le x_{2_0}$ which contradicts our assumption.
We thus conclude that $x_{1_0} = x_{2_0}$.

Combining $x_{1_0}=x_{2_0}=x$ with the two inequalities derived in the first part of this proof, we have the following bounds on $x$.
 \begin{equation}\label{ineqsystem}
    \left\{
                \begin{array}{ll}
                 
                  x<\frac{1}{4}(\frac{v_1}{c_1}+\sqrt{(\frac{v_1}{c_1})^2+4\frac{Rx}{c_1}})\\
                 x<\frac{1}{4}(\frac{v_2}{c_2}+\sqrt{(\frac{v_2}{c_2})^2+4\frac{Rx}{c_2}})
                \end{array}
              \right.
  \end{equation}

We now derive a closed-form solution for the constraint on $x$.
To solve the Inequality System~\ref{ineqsystem}, we write $X=\sqrt{(\frac{v_1}{c_1})^2+4\frac{Rx}{c_1}}$ and, for ease of notation, write $c=c_1$ and $v=v_1$.
Squaring and expanding $X$, one finds that $4x = \frac{c}{R}(X^2-(\frac{v}{c})^2)$.
Accordingly, rewriting the Inequality System~\ref{ineqsystem} in terms of X gives us the following.

 \begin{equation}\label{subineq}
 \frac{c}{R}X^2-X-\frac{v^2}{Rc}-\frac{v}{c}<0 
\end{equation}

With $\Delta'=1+4\frac{c}{R}(\frac{v^2}{Rc}+\frac{v}{c})$, we have that $\Delta'>0$ so the inequality can be simplified as $(X-X_{-}')(X-X_{+}')<0$ with $X_\pm '=R\frac{1\pm \sqrt{\Delta'}}{2c}$.
Since $X_{-}'<X'_{+}$, the solution to Inequality~\ref{subineq} is $X_{-}'<X<X_{+}'$.
As $X'_{-}>0$, we have $(X_{-}')^2<X^2<(X_{+}')^2$ and, therefore, the solutions to the Inequality System~\ref{ineqsystem} are the following, with $\Delta'_1=1+4\frac{c_1}{R}(\frac{v_1^2}{Rc_1}+\frac{v_1}{c_1})$ and $\Delta'_2=1+4\frac{c_2}{R}(\frac{v_2^2}{Rc_2}+\frac{v_2}{c_2})$.

\begin{equation}
    \left\{
                \begin{array}{ll}
                 
                 \frac{c_1}{4R}((R\frac{1-\sqrt{\Delta'_1}}{2c_1})^2-(\frac{v_1}{c_1})^2)<
                 x<\frac{c_1}{4R}((R\frac{1+\sqrt{\Delta'_1}}{2c_1})^2-(\frac{v_1}{c_1})^2)\\
                 \frac{c_2}{4R}((R\frac{1-\sqrt{\Delta'_2}}{2c_2})^2-(\frac{v_2}{c_2})^2)<
                 x<\frac{c_2}{4R}((R\frac{1+\sqrt{\Delta'_2}}{2c_2})^2-(\frac{v_2}{c_2})^2)                 
                \end{array}
              \right.
  \end{equation}

\section{Proof of Theorem~\ref{thm:robustness-new-player}}\label{proof:robustness-new-player}

Consider the strategy of player $n+1$ (the new player) as outlined in~\ref{eq:multiplayer-bestresponse}.
In the 1-value equilibrium we have $\sum_{i=1}^n x_i=n\xeq$.

\begin{enumerate}
\item If $n\xeq\ge \beta$, player $n+1$ is not incentivised to contribute anything so the equilibrium stays unchanged.
\item If $n\xeq<\beta$, there are two cases to consider.
\begin{enumerate}
\item If $\xeq<\beta-n\xeq$ then player $n+1$ contributes $\xeq$.
Since $n\xeq<\beta_j$ (from~\ref{eq:multiplayerNE}) this means that for every player $1\le j\le n$ we are in case (1) of their strategy.
They only change their contribution after the introduction of player $n+1$ if $\beta_j- n\xeq<\xeq$ or equivalently $\beta_j<(n+1)\xeq$.
Since $\beta_1\le \beta_j$, there is at least one player who will change their contribution in this case if and only if $\beta_1<(n+1)\xeq$, in which case we also have $\beta_1<\beta$ from $\xeq<\beta-n\xeq$.
\item If $\xeq\ge\beta-n\xeq$ then player $n+1$ contributes $\beta-n\xeq$.
As before, player $j$ only changes their contribution after the introduction of player $n+1$ if $\beta_j- (n-1)\xeq - (\beta-n\xeq)<\xeq$ or equivalently if $\beta_j<\beta$.
This happens only if $\beta_1<\beta$.
In this case, we also have $\beta_1<\beta<(n+1)\xeq$ from $\xeq\ge\beta-n\xeq$.
\end{enumerate}
\end{enumerate}
A necessary and sufficient condition to having one player changing their contribution is, therefore, $\beta>\beta_1$ and $\frac{1}{n+1}\beta_1<\xeq<\frac{1}{n}\beta$.

\section{Proof of Theorem~\ref{thm:newplayer2value}}\label{proof:newplayer2value}
From Equation~\ref{eq:2-value-eq-cond2}, we have $\sum_{i=1}^n x_i = \beta_1$.
For any new player $n+1$ playing $x_{n+1}>0$ we will, therefore, have  $\sum_{i=1}^{n+1} x_i > \beta_1$, so according to~\ref{eq:multiplayer-bestresponse} the best strategy for player 1 would now be to free ride.
According to~\ref{eq:multiplayer-bestresponse}, the new player contributes if and only if $\sum_{i=1}^n x_i <\beta$ or equivalently $\beta_1<\beta$.
Under this condition, a new player disrupts the equilibrium, which causes player $1$ to free ride.

\section{Proof of Theorem~\ref{thm:player-leaves}}\label{proof:player-leaves}
If a player leaves the game by no longer contributing, we have the following.
In the $1-$value equilibrium case,
since $\xeq\le \frac{1}{n-1}\beta_1<\frac{1}{n-1}\beta_2<\ldots<\frac{1}{n-1}\beta_n$ (implied by Condition~\ref{eq:multiplayerNE}), a player leaving the game does not change the equilibrium (as specified by~\ref{eq:multiplayer-bestresponse}).
Hence a player leaving the game does not disrupt the equilibrium.

\section{Proof of Theorem~\ref{thm:player-leaves-2}}\label{proof:player-leaves-2}
Assume that the player leaving is player $k>1$.
In the case where players are in a $2-$value equilibrium, according to condition~\ref{eq:multiplayer-bestresponse}, player $1$ should change its value to either $\beta_1 - (n-2)\xeq $ or to $\xeq$ or to zero.
Thus, player 1 changes their contribution.

In the case where player 1 is leaving (i.e., $k=1$), then the other players change their contribution if and only if there exists a $j$ such that
$\beta_j-(n-2)\xeq<\xeq$, or  $\beta_j \le(n-1)\xeq $ according to Condition~\ref{eq:multiplayer-bestresponse}.
However due to condition~\ref{eq:2-value-eq-cond1}, such a $j$ does not exist.

\section{Proof of Theorem~\ref{thm:deviation1}}\label{proof:deviation1}
By assumption, every player (other than player $k$) is playing $\xeq$.
After the deviation happens, but before any other changes, players $k$ is playing $x_{k_0}$ and the $n-2$ other players are still contributing $\xeq$.
Each player's new best response then becomes $x_{i,\text{\new}}=\max\{ 0,\min(\max(\xeq,x_{k_0})),(\beta_i-x_{k_0}) - (n-2)\xeq)\}$ due to Condition~\ref{eq:multiplayer-bestresponse}.

Consider the following cases.
\begin{enumerate}
  \item If $x_{k_0}<\xeq$ then $\max(\xeq,x_{k_0})=\xeq$. For each player $i$ we have $x_{k_0}+(n-1)\xeq\le n \xeq\le \beta_i$ and thus $(\beta_i-x_{k_0}) - (n-2)\xeq\ge \xeq$. ($\sum_{j\ne i}x_j<\beta_j$ still holds with $x_{k_0}<\xeq$.)
  This means each player's best response is not impacted.
  Therefore, $\xeq$ is still the equilibrium.

  \item If $x_{k_0}>\xeq$ then we have that $\max(\xeq,x_{k_0})  = x_{k_0}$.
  Therefore, Condition~\ref{eq:multiplayer-bestresponse} becomes $\max\{ 0,\min(x_{k_0},\beta_i- x_{k_0}-(n-2)\xeq)\}$ for each player and for each player $i\ne k_0$ we have the following.
  \begin{enumerate}
    \item If $x_{k_0} \ge \beta_i- x_{k_0}-(n-2)\xeq$, or equivalently $x_{k_0} \ge \frac{\beta_i - (n-2)\xeq}{2}$, player $i$'s new best response is to play $x_{i,\new}=\beta_i - x_{k_0} - (n-2)\xeq$ or 0 if this value is negative.
    This disrupts the current equilibrium and all the rational players have to change their contribution accordingly. 

    \item If $x_{k_0} \le \beta_i- x_{k_0}-(n-2)\xeq$, or equivalently $x_{k_0} \le \frac{\beta_i - (n-2)\xeq}{2}$, player $i$'s best response is to contribute $x_{k_0}$.
    Again, the equilibrium is disrupted.
  \end{enumerate}
\end{enumerate}

To summarise, if a player deviates from the equilibrium and its new contribution is $x_{k_0} \le \xeq$, the best responses of other players stay unchanged and the equilibrium is not disrupted.
In the other case, if $x_{k_0}>\xeq$, the equilibrium changes.

\section{Proof of Theorem~\ref{thm:deviation2}}\label{proof:deviation2}
Assume that $k>1$.
In the case where players are in a $2-$value equilibrium, according to condition~\ref{eq:multiplayer-bestresponse}, player $1$ should change its value to either $\beta_1 - (n-2)\xeq - x_{k_0}$ or to $\max(\xeq,x_{k_0})$ or to zero.
Thus, player 1 changes their contribution.

In the case where player 1 is the deviating player (i.e., $k=1$), then the other players change their contribution if and only if there exists a $j$ such that
$\beta_j-(n-2)\xeq-x_{k_0}<\max(\xeq,x_{k_0})$, or  $\beta_j \le(n-2)\xeq-\max(\xeq,x_{k_0}) $ according to Condition~\ref{eq:multiplayer-bestresponse}.

\end{document}